\documentclass[10pt,letterpaper]{article}
\usepackage{opex3}
\usepackage{color}
\usepackage{graphicx}
\usepackage{amsmath}
\usepackage{amsfonts}
\usepackage{amssymb}

\begin{document}

%% START HERE
%%%%%%%%%%%%%%%%%% title page information %%%%%%%%%%%%%%%%%%
\title{Phase-controlled integrated photonic quantum circuits}

\author{Brian J. Smith$^{1,2,*}$, Dmytro Kundys$^{3}$, Nicholas Thomas-Peter$^2$, P. G. R. Smith$^3$, and I. A. Walmsley$^2$}

\address{$^1$Centre for Quantum Technologies, National University of Singapore\\
3 Science Drive 2, 117543  Singapore, Singapore\\
$^2$University of Oxford, Clarendon Laboratory, Parks Road, \\
Oxford, OX1 3PU, United Kingdom \\ 
$^3$Optoelectronics Research Centre, University of Southampton, Highfield\\
Southampton, SO17 1BJ, United Kingdom}

\vskip-.3cm \parskip0pc\hskip2.25pc \footnotesize 
   \parbox{.8\textwidth}{\begin{center}\it*Corresponding author: \it\textcolor{blue}{\underline{b.smith1@physics.ox.ac.uk}} \rm \end{center} } \normalsize  \vskip-.2cm
% Hacked the opex3.sty email command to get 'corresponding author' in black.
%\email{opex@osa.org} %% email address is required



\begin{abstract}
Scalable photonic quantum technologies are based on multiple nested interferometers. To realize this architecture, integrated optical structures are needed to ensure stable, controllable, and repeatable operation. Here we show a key proof-of-principle demonstration of an externally-controlled photonic quantum circuit based upon UV-written waveguide technology. In particular, we present non-classical interference of photon pairs in a Mach-Zehnder interferometer constructed with X couplers in an integrated optical circuit with a thermo-optic phase shifter in one of the interferometer arms.
\end{abstract}

%\bibliography{Bibliography}
%\bibliographystyle{osajnl}

%%%%%%%%%%%%%%%%%%%%%%% References %%%%%%%%%%%%%%%%%%%%%%%%%%%

%%%%%%%%%%%%%%%%%%%%%%%%%%  body  %%%%%%%%%%%%%%%%%%%%%%%%%%

\section{Introduction}

Quantum-enhanced technologies promise to surpass the performance of their classical counterparts by utilizing inherently quantum-mechanical phenomena, for example, multi-partite entanglement. Certain tasks naturally lend themselves to optical implementations -- small-scale quantum computing and quantum repeaters \cite{kok07, obrien07, DLCZ, gisin07, kimble08}, quantum cryptography \cite{wiesner83, bb84, ekert91, gisin02}, and precision measurement protocols \cite{glm04, glm06, dorner09}.  To implement these quantum technologies, linear optical devices such as beam splitters, phase shifters, and interference filters are required \cite{reck94}. Until recently, the only demonstrations of such quantum devices have been performed using bulk table-top optics \cite{pittman03b,obrien03, obrien04, gasparoni04, langford05, kiesel05b, okamoto05, walther05, kiesel05a, lu07, lanyon07, yuan08, lanyon09, mitchell04, walther04, nagata07}, which is not a practical approach to scalable implementations of more powerful quantum protocols. This is due in part to the difficulty associated with alignment and stabilization of bulk optical elements necessary to create and maintain optimal mode matching in several sequential interferometers. The sheer size of bulk devices is also impractical for scalable manipulation of many photons. Thus an integrated optics approach, which requires no alignment, is inherently stable because of the small size and monolithic structure, and allows easy construction of complex interferometric networks on the micro-scale, provides the best way forward in dealing with these obstacles \cite{walmsley05, walmsley08}.

The performance of photonic quantum technologies is largely governed by the quality of both classical interference and non-classical two-photon or Hong-Ou-Mandel (HOM) interference \cite{HOM} at a beam splitter (or coupler in an integrated optical device) and its multi-photon generalizations \cite{KLM}. Such effective photon-photon interactions are critically affected by the spatial, spectral, and polarization mode overlap of the photons, making mode matching a crucial specification for these devices. 

The information transmitted and manipulated in these quantum applications is typically encoded in a two-level quantum system, called a qubit, which for a single photon can be realized by its polarization, spatial, or spectral mode. In integrated optics, ``dual-rail" logic is typically implemented, in which a qubit is realized by the spatial mode of a photon, that is, one of two paths the photon may take in different waveguides. A logical `1' (`0') is represented by the presence (absence) of the single photon in mode $a(b)$ and the absence (presence) of the photon in mode $b (a)$. In Dirac notation this can be expressed as $|1\rangle_L = |1,0\rangle_{a,b}$ and $|0\rangle_L = |0,1\rangle_{a,b}$, where $|j\rangle_L$, $(j=0,1)$ is the logical state and $|m,n\rangle_{a,b}$ is a photon number state of the electromagnetic field with $m (n)$ photons in mode $a (b)$. With this qubit encoding and careful design of a single waveguide circuit consisting of several interconnected interferometers, each having an adjustable relative phase between the two arms, one can envision programming any desired optical circuit by adjustment of the phases on individual interferometers. This architecture would allow implementation of many different quantum algorithms using the same device and not require the manufacture of several additional circuits.

First steps toward integrated photonic quantum circuits were recently reported based on etched planar silica-on-silicon waveguides \cite{inoue05, politi08} and femtosecond-pulse-written waveguides in bulk silica \cite{marshall09}, and optical fiber \cite{note}. %\footnote{Optical-fiber-based quantum logic gates have been recently demontstrated \cite{chen08, clark09}. Such gates will play an important role in quantum communications, but seem limited to a few gates, due to classical phase stability, rather than computational tasks with several cascaded gate operations.} 
The experiments with silica-on-silicon etched waveguides demonstrated that integrated optical devices can be used to perform high-visibility classical \cite{inoue05} and quantum interference \cite{politi08} at the few-photon level. HOM interference at an evanescent coupler with a visibility of $94.8\pm0.5$\%, and implementation of a controlled-NOT (CNOT) quantum logic gate in the logical basis \cite{politi08} show the feasibility of integrated optics for quantum interference. The output of the CNOT gate was not completely characterized due to the inability to perform full state tomography of the output. Complete characterization requires local single-qubit unitary transformations at each output qubit, that is, the ability to switch to a superposition basis for detection. This can be achieved by using a Mach-Zehnder interferometer (MZI) and two phase shifters. In addition, devices constructed from etched waveguides have considerable loss due to the roughness associated with the etching process, %($\approx$ {\textcolor{red}{xx}}\%), 
and suffer from an inability to cross waveguides over one another, restricting the topology of quantum circuits that can be realized. %{\textcolor{red}{The nature of evanescent couplers (also called directional couplers) used in such devices only allows for narrow band operation ($\approx$ 20 nm at 800 nm central wavelength) over which the splitting ratio is approximately constant ($\eta \approx 0.5 \pm 0.01$).}} 
The femtosecond-laser-writing approach removes the topological restrictions on circuits that can be constructed by writing 3D structures and the loss caused by etching. The crucial ability to control the optical phase between different waveguides and observe high-visibility quantum interference at the same time has yet to be demonstrated in either approach.

In this letter, we present the first quantum interference in phase-controlled integrated photonic quantum circuits. The waveguides are produced by direct UV-writing (DUVW) in planar silica-on-silicon waveguides \cite{svalgaard94}. The UV-writing process enables creation of couplers with varying splitting ratios %{\textcolor{red}{with a wider bandwidth than traditional evanescent couplers}} 
by crossing waveguides at different angles \cite{kundys09a} as illustrated in Fig. \ref{fig:WG}. These X couplers may circumvent the planar restriction by crossing waveguides with essentially no coupling \cite{durr05}, allowing the realization of any circuit topology. Photolithography methods cannot be used to construct X couplers because this technique leaves under-etched segments in the crossing area, resulting in effects such as uncontrolled coupling ratios and high scattering losses. 

Phase control is implemented through a thermo-optic phase shifter, that is, via local heating of a waveguide. This is accomplished by passing a small current through a NiCr electrode deposited above the waveguide. The inherent stability and low propagation loss of the integrated devices paired with the ability to control the phase between different optical paths and excellent mode matching allows observation of phase-dependent quantum effects, such as N00N-state interference \cite{rarity90, mitchell04, nagata07}. 

\section{Waveguides}

In traditional integrated optical devices, light is guided along a core channel embedded in a cladding of slightly lower refractive index (similar to an optical fiber). These channels are typically prepared from semiconductor wafers using lithographic etching techniques. Support of a single transverse mode for a given wavelength range is achieved by careful choice of the waveguide dimensions and refractive indices of the core and cladding layers. Coupling between waveguides can be achieved by bringing together the two guides sufficiently close so that their evanescent fields overlap. The coupling ratio $\eta$, which is equivalent to the beam splitter reflectivity, is determined by the distance between the waveguides and the length of the coupler. This evanescent coupler, also known as a directional coupler, operates by an interferometric effect so that the coupling ratio (or effective beam splitter reflectivity) is wavelength dependent. The interferometric nature of the coupler also makes it sensitive to temperature and polarization fluctuations at the coupling junction as well as unavoidable manufacturing imperfections. In previous work we have demonstrated that compact X couplers produced by DUVW can be used to improve the coupler stability through careful refractive index matching in the waveguide crossing area \cite{kundys09a}.

\begin{figure}[h]
\centering\includegraphics[width=0.9\textwidth]{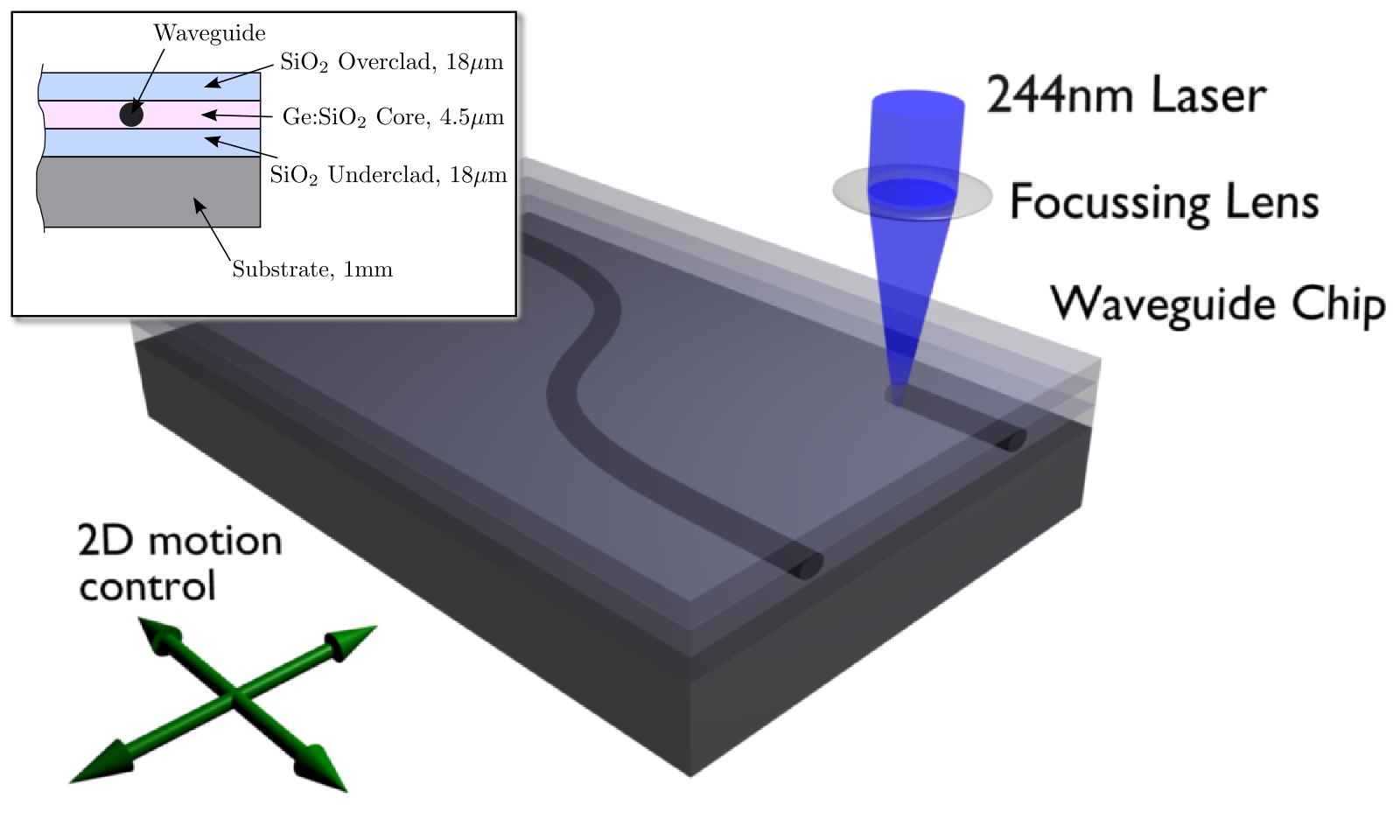}
\caption{Schematic of the UV-waveguide writing process and their structure (inset).}
\label{fig:WG}
\end{figure}

Laser written waveguides allow creation of structures that are not viable with lithographic etching techniques of standard integrated optics manufacturing such as X couplers \cite{kundys09a, adikan06}, and Bragg reflectors \cite{emmerson02}. Since there is no lithographic step, the waveguide writing process is quite rapid and enables quick development from prototype to completed device. %Our waveguides are constructed from a planar silica-on-silicon wafer that is first diced into several chips of typical dimensions 25 mm $\times$ 15 mm (length $\times$ width). The wafer is composed of a silicon layer ({\textcolor{red}{xx}} $\mu$m) for structural integrity with a planar waveguide structure deposited on top via flame-hydrolysis deposition (FHD). 
The host-slab waveguides are fabricated by flame-hydrolysis deposition (FHD) \cite{tandon03}, utilizing 6" diameter silicon wafers (1 mm thick) with a 20 $\mu$m thick thermal-oxide silica layer as a substrate and further deposition of doped silica layers to provide in-plane optical confindement. The wafer is subsequently diced into chips with typical dimensions of 25 mm $\times$ 10 mm (length $\times$ width). The guiding structure consists of a lower silica cladding layer (18 $\mu$m), a guiding germanium-doped ($\approx$ 0.5 \% atomic mass, measured by energy dispersive X-ray (EDX) spectroscopy) silica core (4.5 $\mu$m), and a top silica cladding layer (18 $\mu$m). The individual waveguides are written by focusing a continuous-wave UV laser (244 nm wavelength) onto the chip which is subsequently moved transversely to the surface normal with computer-interfaced 2D motion control, as depicted in Fig. \ref{fig:WG}. The core of the waveguide is formed by a local change in refractive index of the Ge-doped planar silica layer that is proportional to the local fluence of the UV light, and is roughly $\delta n \approx 4 \times 10^{-3}$ \cite{kundys09a, adikan06}. This leads to an effective step-index profile for the guiding core. To ensure symmetric single transverse spatial mode operation of the waveguides at 800 nm wavelength (where high-efficiency single-photon detectors are available), care must be taken in choosing the core dimension, laser focusing and fluence \cite{kundys09a, adikan06}. The UV laser was focused to a spot size of approximately 4.5 $\mu$m to match the core thickness and ensure a circular guiding mode. The cladding layers were doped with Boron to closely match the index of the Ge-doped core layer. This implies that the photo-induced index change is the main contribution to optical guiding.  The supported transverse waveguide mode is extremely well matched to single-mode optical fibers, with measured butt-coupling efficiencies of up to 70\% in a single channel. In our experiments, we operate with  $\approx$60\% coupling efficiency at the inputs and outputs due to difficulties coupling multiple channels at the same time.

The effective beam splitter reflectivity of the X couplers is primarily governed by the intersection angle of the guides (2.45$^{\circ}$ for 50:50 coupling) and the balance of the UV-light fluence (and thus the index change $\delta n$) between both waveguides at the crossing region. The X couplers are formed by writing one channel at a time taking care to decrease the laser flux at the crossing point to ensure a continuous refractive index profile, as depicted in Fig. \ref{fig:WG}. Due to the dependence of the coupling ratio on geometry as opposed to optical interference, which occurs in evanescent couplers, X couplers are less sensitive to fabrication imperfections, thermal fluctuations, and polarization state at the coupling junction \cite{kundys09a, adikan06}. % {\textcolor{red}{Due to the dependence of the coupling ratio on geometry as opposed to optical interference, the X couplers exhibit nearly constant splitting ratio ($\leq$ {\textcolor{red}{xx}}\% difference) over a broad spectral range ({\textcolor{red}{xx}} nm bandwidth at {\textcolor{red}{xx}} nm wavelength) and are not sensitive to thermal fluctuations or polarization state at the coupling junction \cite{kundys09a, adikan06}.}}

To control the phase between two different waveguides on the chip a thermo-optic phase shifter is used. A small NiCr electrode (0.4 $\mu$m $\times$ 50 $\mu$m $\times$ 2.5 mm = height $\times$ width $\times$ length, 0.85 k$\Omega$ electrical resistance) is deposited directly over one of the waveguides through which a current can be passed. The local Ohmic heating causes the refractive index to change slightly, thus inducing an optical phase difference between the two different waveguides. This phase can be adjusted across a full range from 0 to 2$\pi$. %In operation, thermo-optic elements cause slight sample heating ($\approx$ 1 Watt), which heats the sample. This results in some tuning transients that stabilize within a few minutes {\textcolor{red}{[Chris paper?]}}.
%{\textcolor{red}{and it is repeatable to within {\textcolor{red}{xx}}\%. The phase is stable over a time period much longer than our data acquisition ({\textcolor{red}{xx}} hours - minutes - seconds ???). We anticipate that the performance of the thermo-optic phase shifters can be improved by thermally isolating the chip on a thermo-electric cooler to extract any excess heat. The time response of the thermo-optic  phase shifter is approximately {\textcolor{red}{xx}} radians / ms.}

\section{Experiment}

\begin{figure}[h]
\centering\includegraphics[width=0.9\textwidth]{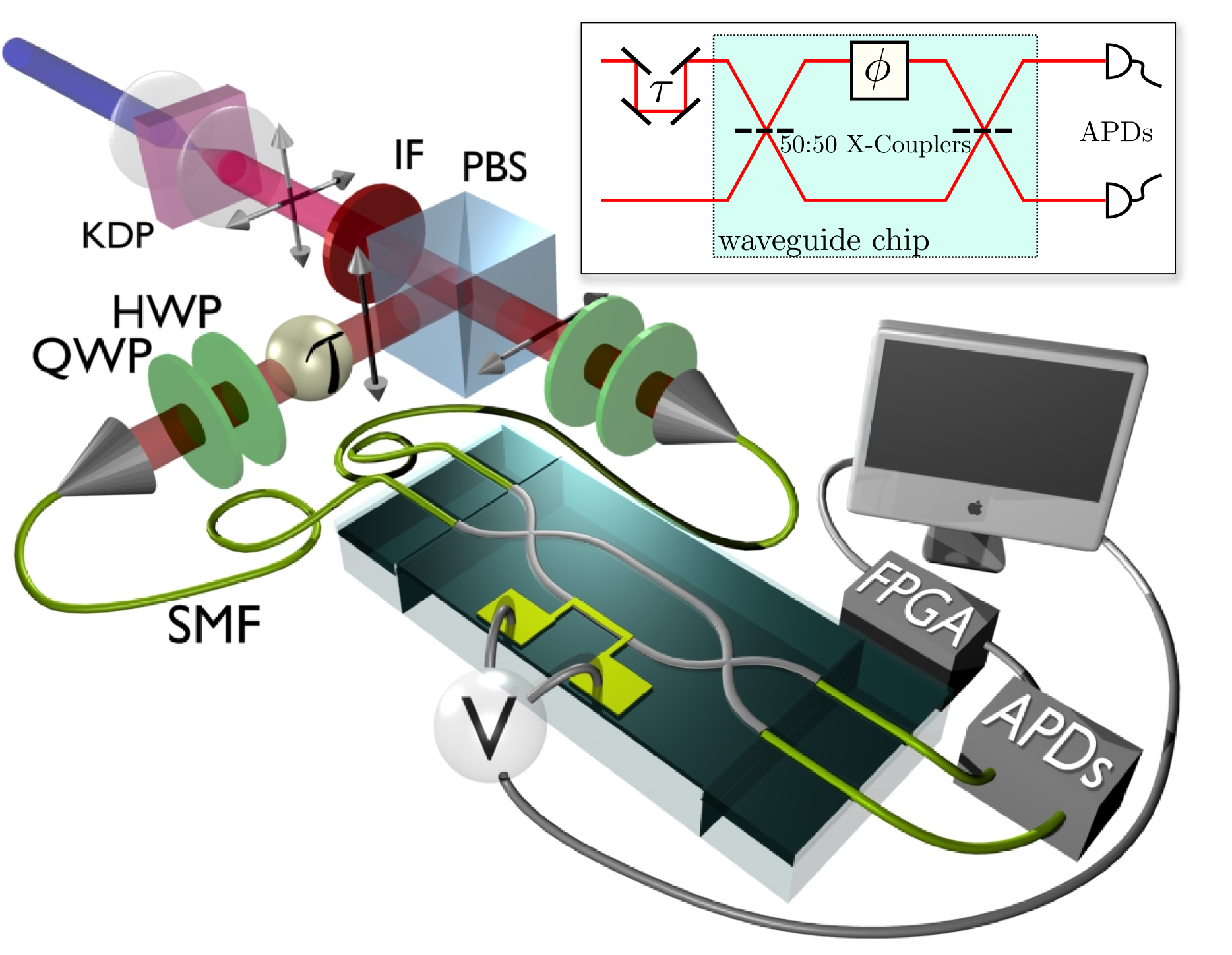}
\caption{Experimental setup. Horizontally and vertically polarized photon pairs created by SPDC pass through an interference filter (IF) prior to being separated by a polarizing beam splitter (PBS). A temporal delay line ($\tau$) in one of the photon paths controls the relative arrival time of the photons to the MZI. Half and quarter wave plate pairs (HWP and QWP) ensure the photons coupled into single-mode fibers (SMFs) arrive at the waveguide with identical polarization states. A computer-controlled voltage supply (V) controls the relative phase, $\phi$, between the interferometer arms. The outputs of the interferometer are sent to Si avalanche photodiodes (APDs). The singles and coincidence count rates are monitored by field-programmable gate array (FPGA) electronics connected to a computer. Inset shows a schematic of the setup.}
\label{fig:expsetup}
\end{figure}

To demonstrate the capability of our UV-written waveguides in the quantum regime, we performed two experiments using a Mach-Zehnder interferometer (MZI) as shown in Fig. \ref{fig:expsetup}. First we demonstrate the mode-matching capabilities and phase stability of the integrated optics and X-coupler performance by observing non-classical two-photon Hong-Ou-Mandel (HOM) interference \cite{HOM}. Here the MZI is used as a beam splitter with effective reflectivity $\eta = (1+\cos(\Delta \phi))/2$. The phase difference between the two interferometer arms is $\Delta \phi = \phi - \phi_0$, with $\phi$ the phase change due to the thermo-optic phase shifter, and $\phi_0$ any initial phase difference between the arms when no current is applied. 

In HOM interference two photons are incident at opposite input ports of a 50:50 beam splitter. If the photons are identical in all degrees of freedom (polarization, spatial, and spectral including phase), then the photons will always exit the beam splitter in the same output port. This can be seen as a consequence of the bosonic nature of photons. If some distinguishability between the photons is introduced, say by changing their relative arrival time to the beam splitter, then there is a chance that the photons will exit the beam splitter from different ports. By placing photon-counting detectors at the output ports of this beam splitter and examining the coincidence statistics for repeated experiments as a function of the relative time delay between the photons, the characteristic HOM dip in coincidence rate at zero time delay is observed \cite{HOM}. The visibility of the HOM dip sets a lower bound on the photon indistinguishability and thus mode overlap \cite{mosley08a, mosley08b}.

The photons used in our experiments, centered at 830 nm wavelength, are created from a degenerate, collinear, type-II spontaneous parametric downconversion (SPDC) source  \cite{mosley08a, mosley08b}. The 5 mm long potassium-diphosphate (KDP) crystal is pumped with 600 mW of frequency-doubled pulses (415 nm wavelength and 3.5 nm bandwidth) derived from a Ti:Sapphire laser. The downconverted light is passed through an interference filter (IF) (830 nm central wavelength and 3 nm bandwidth) to ensure good spectral overlap of the photons, prior to separating the two polarization modes on a polarizing beam splitter (PBS). Quarter and half wave plates (QWP and HWP) are used prior to launching each polarization state into single-mode fibers (SMFs) to ensure that each beam propagates through the MZI with the same polarization. A temporal delay line ($\tau$) in the path of one of the photons allows control of the relative arrival time of the photons at the MZI. The SMFs were directly coupled to the waveguides using a v-groove assembly (Oz Optics) and fixed by index-matched epoxy. The spacing of the waveguides (250 $\mu$m) was chosen to match the commercially available fiber array used in the experiments. 

\begin{figure}[h]
\centering\includegraphics[width=0.9\textwidth]{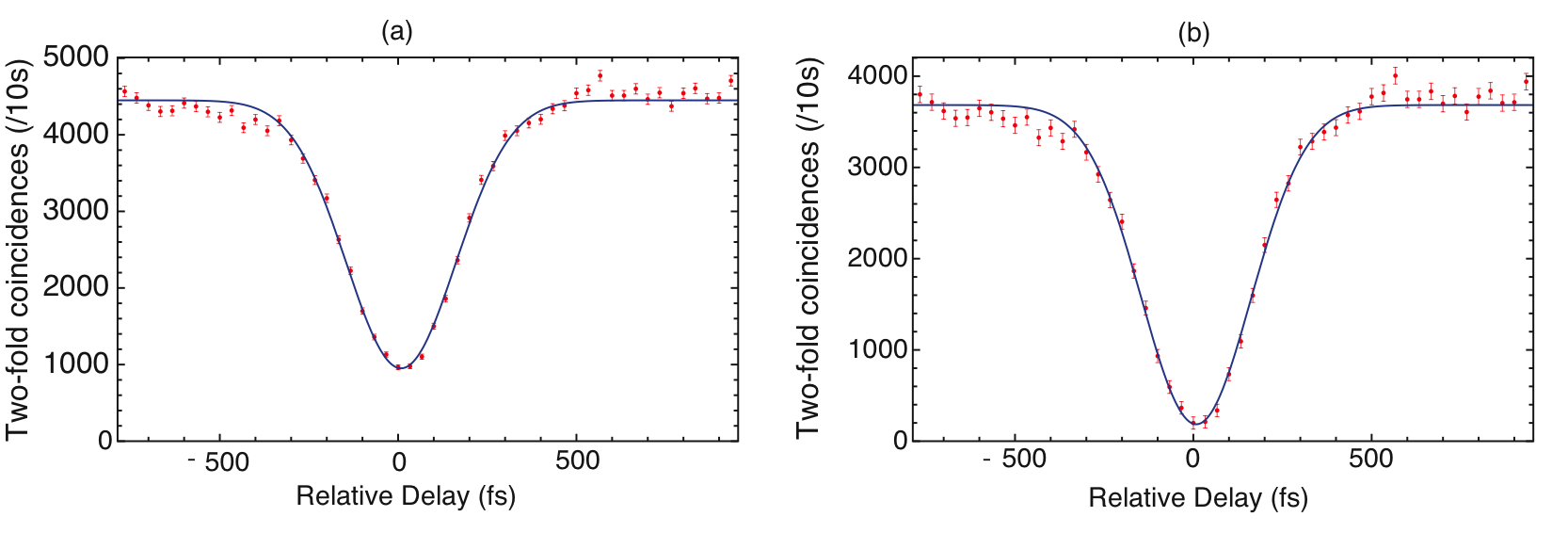}
\caption{Two-photon HOM interference through a waveguide MZI with the phase ($\phi$) tuned to act as a 50:50 beam splitter. Plots show (a) measured coincidence count rate ($79 \pm 1$\% visibility) and (b) background subtracted coincidence count rate ($95.0 \pm 1.4$\% visibility) as a function of the relative arrival time of photons to the beam splitter, $\tau$. Error bars are calculated assuming Poisson count statistics and propagating errors for the background subtraction. The blue line is a Gaussian fit.}
\label{fig:HOMI}
\end{figure}

Setting the relative optical phase in the MZI so that it performed as a nearly 50:50 beam splitter and delaying the arrival time of the photons by adjusting the optical delay line, we obtain the two-photon HOM interference dip shown in Fig. \ref{fig:HOMI} (a). The raw visibility of the dip ($79 \pm 1$\%), defined by $V=(R_{\rm{max}}-R_{\rm{min}})/R_{\rm{max}}$ and obtained from a Gaussian fit, is at first not so encouraging, since much better interference has been observed in bulk optics  \cite{mosley08a, mosley08b}. However, this is mainly due to significant contamination arising from more than one photon pair being produced in the SPDC source. This background could be reduced by lowering the SPDC pump power, but this would significantly increase the data acquisition time. Since we are interested mainly in the device characteristics and not the source specifications, we subtract the effects of the background. To determine this background we block one of the MZI inputs and count the number of coincidences in 300 seconds. This is repeated with the other input blocked and the sum normalized by the counting time is taken as the background ($\approx  75$ / s) which is subtracted from the raw data. The resulting background-subtracted dip is shown in Fig. \ref{fig:HOMI} (b), which gives improved visibility ($95.0 \pm 1.4$\%), indicating better mode overlap. This demonstrates the ability to set a stable phase over the measurement time ($\approx 10$ min.) and observe high-visibility quantum interference.

The residual 5\% visibility is attributed to the polarization mode mismatch of the input photons. Even though the waveguide X couplers are polarization insensitive, to observe high-quality quantum interference requires that the photons are launched into the guides with identical polarizations, which is difficult to achieve with the non-polarization-preserving single-mode-fiber v-groove assemblies used here. This can be addressed by using polarization-maintaining-fiber v grooves in future applications. 

To show the phase-control performance of our integrated optical circuits at the quantum level, we perform a two-photon N00N interference experiment. In an interferometer, N00N states of light lead to a fringe pattern in N-fold coincidence photodetection events that oscillates N-times faster as a function of the phase than the classical fringe pattern. A N00N state of the electromagnetic field is one in which two modes $a(b)$ are in an equal superposition of having N(0) photons in mode $a(b)$ and 0(N) photons in mode $a(b)$, i.e. the state of the field is $|\psi\rangle = (|{\rm{N}},0\rangle + |0,{\rm{N}}\rangle)/\sqrt{2}$. Here $|m,n\rangle$ is a Fock state with $m(n)$ photons in mode $a(b)$. We can realize a two-photon N00N state inside the MZI by launching the state $|1,1\rangle$ into the inputs. This state transforms after the first coupler into the superposition
\begin{equation}
|1,1\rangle \rightarrow ( |2,0\rangle - |0,2\rangle ) / \sqrt{2},
\label{eq:1}
\end{equation}
due to the HOM effect. The right-hand side of Eq. (\ref{eq:1}) is the two-photon N00N state. If there is a phase shift, $\phi$, in mode $a$ relative to $b$, the state evolves as
\begin{equation}
( |2,0\rangle - |0,2\rangle ) / \sqrt{2} \rightarrow ( |2,0\rangle e^{i 2 \phi} - |0,2\rangle ) / \sqrt{2},
\label{eq:3}
\end{equation}
where the two-photon component in mode $a$ picks up twice the phase  $\phi$. After the second beam splitter we project onto the state $|1,1\rangle$, that is, collect the two-fold coincidence counts at the output of the MZI. This leads to a coincidence rate as a function of the relative phase $\phi$ given by
\begin{equation}
R_{1,1}(\phi) = R_0\left\{1 + A \cos \left[2( \phi+\phi_0)\right]\right\}.
\label{eq:5}
\end{equation}
Here $R_0$ is an overall count rate, $A=(R_{\rm{max}}-R_{\rm{min}})/(R_{\rm{max}}+R_{\rm{min}})$ is the fringe visibility, and $\phi_0$ is the initial phase difference in the interferometer. Hence the characteristic 2$\phi$ fringe pattern in the coincidence rate for the two-photon N00N state can be observed. The visibility of these fringes is a measure of the indistinguishability of the photons propagating in the waveguides, and thus characterizes the performance of the couplers, phase shifter, and waveguides at the quantum level.% For imperfect mode matching, $O$, the two-fold coincidence probability of a two-photon N00N state is given by
%\begin{equation}
%P_{1,1}(\phi,O) = \frac{1}{2}\left[\left(\frac{3-O}{2}\right)+\left(\frac{1+O}{2}\right)\cos(2\phi)\right].
%\label{eq:6}
%\end{equation}
%The visibility, $V$, of the two-fold fringe pattern in Eq. (\ref{eq:6}) is related to the mode overlap of the two photons by $O = (3V - 1) / (1 + V)$.

\begin{figure}[h]
\centering\includegraphics[width=0.9\textwidth]{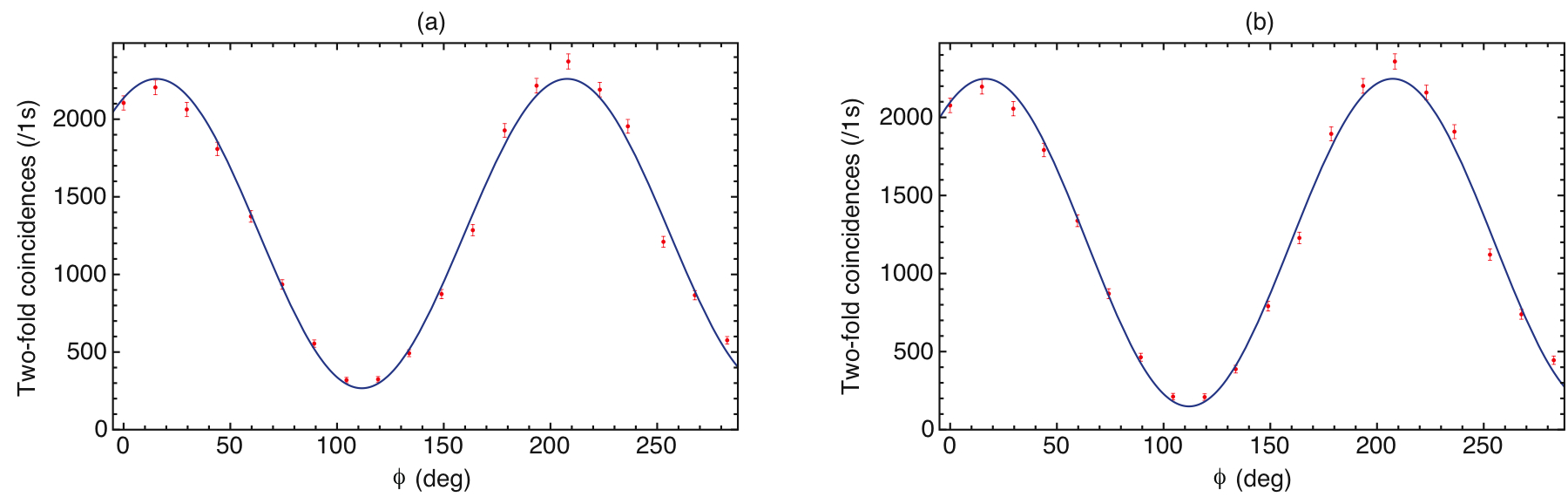}
\caption{Two-photon N00N interference through a waveguide MZI as a function of the phase induced by the thermo-optic phase shifter. Plots show (a) raw measured coincidence count rate ($78.9 \pm 3$\% visibility) and (b) background subtracted coincidence count rate ($88 \pm 3$\% visibility) at zero time delay. Error bars are calculated assuming Poisson count statistics and propagating errors for the background subtraction. The blue line is a fit to Eq. (\ref{eq:5}).}
\label{fig:N00N}
\end{figure}

The two-photon N00N state is produced from the same SPDC source used in the two-photon HOM interference experiment, with the time delay between the beams set to zero. The raw two-fold coincidence rates as a function of the phase induced by the thermo-optic phase shifter, $\phi$, is plotted in Fig. \ref{fig:N00N} (a). The phase is related to the dissipated power, P, by $\phi = \alpha {\rm{P}}$. Here $\alpha = 0.579 \pm 0.006$ deg/mW is determined from the classical fringe pattern. The power is given by ${\rm{P}} = V^2/R$, where $V$ is the applied voltage and $R = 850 \pm 5$ $\Omega$ is the measured resistance of the thermo-optic phase shifter.

Again we note that there is a significant background coincidence contribution from multiple photon events similar to those that occurred in the HOM interference experiment. This can be taken into account by measuring the two-fold coincidence count rates as the phase is varied for each of the SPDC modes blocked. These background count rates can then be subtracted from the raw data giving a fringe visibility of $88 \pm 3$\% as shown in Fig. \ref{fig:N00N} (b). The visibility is not equal to the 95\% observed in the HOM interference because the waveguide used for the HOM experiment was damaged by driving the thermo-optic phase shifter with too much power. 

The discrepancy in the visibilities of the HOM (95\%) and N00N-state (88\%) could in principle be due to the phase shifter introducing distinguishability between the modes, the waveguides may be polarization sensitive, or input photon mode mismatch. We can rule out the first two of these since they would affect HOM interference visibility in all waveguides. The 95\% visibility observed in the HOM experiment taken over 10 minutes quantifies the minimum performance of the waveguide circuits. This sets a quantitative lower bound on their mode matching and stability capabilities. However, we are not able to rule out the input state of light. The polarization mode mismatch of the photons launched into the waveguide is not trivial to determine. This can be addressed by using polarization-preserving optical fibers to couple light into the waveguides.

\section{Conclusions}

Integrated optics is a promising route to scalable quantum technologies. The UV-written waveguide circuits demonstrated here have several advantages over structures developed with lithographic techniques. The ability to create X couplers as opposed to evanescent (directional) couplers, which operate over a larger bandwidth, can be used to circumventing the topological restrictions of the planar 2D structure and enables new circuit designs. The laser writing process allows for rapid development of circuit layouts and can be used to create other devices such as Bragg reflectors in situ. We have shown the first phase-controlled photonic quantum circuits in an integrated optical device, which is a key requirement for future developments in quantum technologies. The observed 95\% HOM interference visibility and 88\% two-photon N00N-state fringe visibility are promising indicators that controllable quantum circuits of nested interferometers can be constructed. Although further development of the polarization invariance of the X couplers is needed. In addition, this technology sets the stage for development of more complicated quantum circuits and applications, such as a quantum thermal sensor. 

Note added in proof: We note that similar results to those reported here have recently been published \cite{matthews09}.

\section*{Acknowledgments}
This work was supported by the EPSRC through the QIP IRC and grant EP/C013840/1, the EC under the Integrated Project QAP, and the Royal Society.


\begin{thebibliography}{60}
\newcommand{\enquote}[1]{``#1''}
\expandafter\ifx\csname url\endcsname\relax
  \def\url#1{\texttt{#1}}\fi
\expandafter\ifx\csname urlprefix\endcsname\relax\def\urlprefix{URL }\fi
\providecommand{\eprint}[2][]{\url{#2}}

\bibitem{kok07}
P.~Kok, W.~J. Munro, K.~Nemoto, T.~C. Ralph, J.~P. Dowling, and G.~J. Milburn,
  \enquote{Linear optical quantum computing with photonic qubits,} Rev. Mod.
  Phys. \textbf{79}, 135--174 (2007).

\bibitem{obrien07}
J.~L. O'Brien, \enquote{Optical Quantum Computing,} Science \textbf{318},
  1567--1570 (2007).

\bibitem{DLCZ}
L.~M. Duan, M.~D. Lukin, J.~I. Cirac, and P.~Zoller, \enquote{Long-distance
  quantum communciation with atomic ensembles and linear optics,} Nature
  \textbf{414}, 413--418 (2001).

\bibitem{gisin07}
N.~Gisin and R.~Thew, \enquote{Quantum communication,} Nat. Photon.
  \textbf{1}, 165--171 (2007).

\bibitem{kimble08}
H.~J. Kimble, \enquote{The quantum internet,} Nature \textbf{453}, 1023--1030
  (2008).

\bibitem{wiesner83}
S.~Wiesner, \enquote{Conjugate coding,} SIGACT News \textbf{15}, 78--88 (1983).

\bibitem{bb84}
C.~H. Bennett and G.~Brassard, \enquote{Quantum Cryptography: Public Key Distribution and Coin Tossing,} in \emph{Proceedings of the IEEE International Conference on Computers, Systems and Signal Processing},  (IEEE, 1984) pp. 175--179.

\bibitem{ekert91}
A.~K. Ekert, \enquote{Quantum cryptography based on Bell's theorem,} Phys. Rev.
  Lett. \textbf{67}, 661--663 (1991).

\bibitem{gisin02}
N.~Gisin, G.~Ribordy, W.~Tittel, and H.~Zbinden, \enquote{Quantum
  cryptography,} Rev. Mod. Phys. \textbf{74}, 145--195 (2002).

\bibitem{glm04}
V.~Giovannetti, S.~Lloyd, and L.~Maccone, \enquote{Quantum-Enhanced
  Measurements: Beating the Standard Quantum Limit,} Science \textbf{306},
  1330--1336 (2004).

\bibitem{glm06}
V.~Giovannetti, S.~Lloyd, and L.~Maccone, \enquote{Quantum Metrology,} Phys.
  Rev. Lett. \textbf{96}, 010401 (2006).

\bibitem{dorner09}
U.~Dorner, R.~Demkowicz-Dobrzanski, B.~J. Smith, J.~S. Lundeen, W.~Wasilewski,
  K.~Banaszek, and I.~A. Walmsley, \enquote{Optimal Quantum Phase Estmation,}
  Phys. Rev. Lett. \textbf{102}, 040403 (2009).

\bibitem{reck94}
M. Reck, A. Zeilinger, H. J. Bernstein, and P. Bertani, \enquote{Experimental Realization of Any Discrete Unitary Operator,}
  Phys. Rev. Lett. \textbf{73}, 58--61 (1994).

\bibitem{pittman03b}
T.~B. Pittman, M.~J. Fitch, B.~C. Jacobs, and J.~D. Franson,
  \enquote{Experimental controlled-NOT logic gate for single photons in the
  coincidence basis,} Phys. Rev. A \textbf{68}, 032316 (2003).

\bibitem{obrien03}
J.~L. O'Brien, G.~J. Pryde, A.~G. White, T.~C. Ralph, and D.~Branning,
  \enquote{Demonstration of an all-optical quantum controlled-NOT gate,} Nature
  \textbf{426}, 264--267 (2003).

\bibitem{obrien04}
J.~O'Brien, G.~J. Pryde, A.~Gilchrist, D.~F.~V. James, N.~K. Langford, T.~C.
  Ralph, and A.~G. White, \enquote{Quantum Process Tomography of a
  Controlled-NOT Gate,} Phys. Rev. Lett. \textbf{93}, 080502 (2004).

\bibitem{gasparoni04}
S.~Gasparoni, J.-W. Pan, P.~Walther, T.~Rudolph, and A.~Zeilinger,
  \enquote{Realization of a Photonic Controlled-NOT Gate Sufficient for Quantum
  Computation,} Phys. Rev. Lett. \textbf{93}, 020504 (2004).

\bibitem{langford05}
N.~K. Langford, T.~J. Weinhold, R.~Prevedel, K.~J. Resch, A.~Gilchrist, J.~L.
  O'Brien, G.~J. Pryde, and A.~G. White, \enquote{Demonstration of a Simple
  Entangling Optical Gate and Its Use in Bell-State Analysis,} Phys. Rev. Lett.
  \textbf{95}, 210504 (2005).

\bibitem{kiesel05b}
N.~Kiesel, C.~Schmid, U.~Weber, R.~Ursin, and H.~Weinfurter, \enquote{Linear
  Optics Controlled-Phase Gate Made Simple,} Phys. Rev. Lett. \textbf{95},
  210505 (2005).

\bibitem{okamoto05}
R.~Okamoto, H.~F. Hofmann, S.~Takeuchi, and K.~Sasaki, \enquote{Demonstration
  of and Optical Quantum Controlled-NOT Gate withtout Path Interference,} Phys.
  Rev. Lett. \textbf{95}, 210506 (2005).

\bibitem{walther05}
P.~Walther, K.~J. Resch, T.~Rudolph, E.~Schenck, H.~Weinfurter, V.~Vedral,
  M.~Aspelmeyer, and A.~Zeilinger, \enquote{Experimental one-way quantum
  computing,} Nature \textbf{434}, 169--176 (2005).

\bibitem{kiesel05a}
N.~Kiesel, C.~Schmid, U.~Weber, G.~Toth, O.~Guhne, R.~Ursin, and H.~Weinfurter,
  \enquote{Experimental Analysis of a Four-Qubit Photon Cluster State,} Phys.
  Rev. Lett. \textbf{95}, 210502 (2005).

\bibitem{lu07}
C.~Y. Lu, X.-Q. Zhou, O.~Guhne, W.-B. Gao, J.~Zhang, Z.-S. Yuan, A.~Goebel,
  T.~Yang, and J.-W. Pan, \enquote{Experimental entanglement of six photons in
  graph states,} Nat. Phys. \textbf{3}, 91--95 (2007).

\bibitem{lanyon07}
B.~P. Lanyon, T.~J. Weinhold, N.~K. Langford, M.~Barbieri, D.~F.~V. James,
  A.~Gilchrist, and A.~G. White, \enquote{Experimental Demonstration of a
  Compiled Version of Shor's Algorithm with Quantum Entanglement,} Phys. Rev.
  Lett. \textbf{99}, 250505 (2007).

\bibitem{yuan08}
Z.-S. Yuan, Y.-A. Chen, B.~Zhao, S.~Chen, J.~Schmiedmayer, and J.-W. Pan,
  \enquote{Experimental demonstration of a BDCZ quantum repeater node,} Nature
  \textbf{454}, 1098--1101 (2008).

\bibitem{lanyon09}
B.~P. Lanyon, M.~Barbieri, M.~P. Almeida, T.~Jennewein, T.~C. Ralph, K.~J.
  Resch, G.~J. Pryde, J.~L. O'Brien, A.~Gilchrist, and A.~G. White,
  \enquote{Simplifying quantum logic using higher-dimensional Hilbert spaces,}
  Nat. Phys. \textbf{5}, 134--140 (2009).

\bibitem{mitchell04}
M.~W. Mitchell, J.~S. Lundeen, and A.~M. Steinberg, \enquote{Super-Resolving
  Phase Measurements with a Multiphoton Entangled State,} Nature \textbf{429},
  161--164 (2004).

\bibitem{walther04}
P.~Walther, J.-W. Pan, M.~Aspelmeyer, R.~Ursin, S.~Gasparoni, and A.~Zeilinger,
  \enquote{De Broglie wavelength of a non-local four-photon state,} Nature
  \textbf{429}, 158--161 (2004).

\bibitem{nagata07}
T.~Nagata, R.~Okamoto, J.~L. O'Brien, K.~Sasaki, and S.~Takeuchi,
  \enquote{Beating the Standard Quantum Limit with Four-Entangled Photons,}
  Science \textbf{316}, 726-729 (2007).

\bibitem{walmsley05}
I.~A. Walmsley and M.~G. Raymer, \enquote{Toward Quantum-Information Processing
  with Photons,} Science \textbf{307}, 1733--1734 (2005).

\bibitem{walmsley08}
I.~A. Walmsley, \enquote{Looking to the Future of Quantum Optics,} Science
  \textbf{319}, 1211--1213 (2008).

\bibitem{HOM}
C.~K. Hong, Z.~Y. Ou, and L.~Mandel, \enquote{Measurement of subpicosecond time
  intervals between two photons by interference,} Phys. Rev. Lett. \textbf{59},
  2044--2046 (1987).

\bibitem{KLM}
R.~Laflamme, G.~J. Milburn, and E.~Knill, \enquote{A scheme for efficient
  quantum computation with linear optics,} Nature \textbf{409}, 46--52 (2001).

\bibitem{inoue05}
H.~Takesue and K.~Inoue, \enquote{Generation of 1.5-$\mu$m band time-bin
  entanglement using spontaneous fiber four-wave mixing and planar light-wave
  circuit interferometers,} Phys. Rev. A \textbf{72}, 041804(R) (2005).

\bibitem{politi08}
A.~Politi, M.~J. Cryan, J.~G. Rarity, S.~Yu, and J.~L. O'Brien,
  \enquote{Silica-on-Silicon Waveguide Quantum Circuits,} Science \textbf{320},
  646--649 (2008).

\bibitem{marshall09}
G.~D. Marshall, A.~Politi, J.~C.~F. Matthews, P.~Dekker, M.~Ams, M.~J.
  Withford, and J.~L. O'Brien, \enquote{Laser written photonic quantum
  circuits,} 	arXiv:0902.4357v1.

\bibitem{note}
Optical-fiber-based quantum logic gates have been recently demonstrated \cite{chen08, clark09}. Such gates will play an important role in quantum communications, but seem limited to a few gates due to classical phase stability, rather than computational tasks with several cascaded gate operations.

\bibitem{chen08}
J.~Chen, J.~B. Altepeter, M.~Medic, K.~F. Lee, B.~Gokden, R.~H. Hadfield, S.~W.
  Nam, and P.~Kumar, \enquote{Demonstration of a Quantum Controlled-NOT Gate in
  the Telecommunications Band,} Phys. Rev. Lett. \textbf{100}, 133603 (2008).

\bibitem{clark09}
A.~S. Clark, J.~Fulconis, J.~G. Rarity, W.~J. Wadsworth, and J.~L. O'Brien,
  \enquote{All-optical-fiber polarization-based quantum logic gate,} Phys. Rev.
  A \textbf{79}, 030303(R) (2009).

\bibitem{svalgaard94}
M.~Svalgaard, C.~V. Poulsen, A.~Bjarklev, and O.~Poulsen, \enquote{UV-writing
  of buried single-mode channel waveguides in Ge-doped silica films,} Electron.
  Lett. \textbf{30}, 1401--1402 (1994).

\bibitem{kundys09a}
D.~O. Kundys, J.~C. Gates, S.~Dasgupta, C.~B.~E. Gawith, and P.~G.~R. Smith,
  \enquote{Use of Cross-Couplers to Decrease Size of UV Written Photonic
  Circuits,} IEEE Photon. Technol. Lett. (to be published).

\bibitem{durr05}
F.~Durr and H.~Renner, \enquote{Analytical Design of X-Couplers,} J. Lightwave
  Technol. \textbf{23}(2), 876--885 (2005).

\bibitem{rarity90}
J.~G. Rarity, P.~R. Tapster, E.~Jakeman, T.~Larchuk, R.~A. Campos, M.~C. Teich,
  and B.~E. Saleh, \enquote{Two-photon interference in a Mach-Zehnder
  interferometer,} Phys. Rev. Lett. \textbf{65}, 1348--1351 (1990).

\bibitem{adikan06}
F.~R.~M. Adikan, C.~B.~E. Gawith, P.~G.~R. Smith, I.~J.~G. Sparrow,
  G.~D.Emmerson, C.~Riziotis, and H.~Ahmad, \enquote{Design and demonstration
  of direct UV-written small angle X couplers in silica-on-silicon for
  broadband operation,} Appl. Opt. \textbf{45}, 6113--6118 (2006).

\bibitem{emmerson02}
G.~D. Emmerson, S.~P. Watt, C.~B.~E. Gawith, V.~Albanis, M.~Ibsen, R.~B.
  Williams, and P.~G.~R. Smith, \enquote{Fabrication of directly UV-written
  channel waveguides with simultaneously defined integral Bragg gratings,}
  Electron. Lett. \textbf{38}, 1531--1532 (2002).

\bibitem{tandon03}
P.~Tandon and H.~Boek, \enquote{Experimental and theoretical studies of flame
  hydrolosis deposition process for making glasses for optical planar devices,}
  J. Non-Crystal. Solids \textbf{317}, 275--289 (2003).

\bibitem{mosley08a}
P.~J. Mosley, J.~S. Lundeen, B.~J. Smith, P.~Wasylcyk, A.~B. U'Ren,
  C.~Silberhorn, and I.~A. Walmsley, \enquote{Heralded Generation of Ultrafast
  Single Photons in Pure Quantum States,} Phys. Rev. Lett. \textbf{100},
  133601 (2008).

\bibitem{mosley08b}
P.~J. Mosley, J.~S. Lundeen, B.~J. Smith, and I.~A. Walmsley,
  \enquote{Conditional preparation of single photons using parametric
  downconversion: a recipe for purity,} New J. Phys. \textbf{10}, 093011
  (2008).

\bibitem{matthews09}
J.~C.~F. Matthews, A. Politi, A. Stefanov, and J.~L. O'Brien,
  \enquote{Manipulation of multiphoton entanglement in waveguide quantum circuits,} Nat. Photon. \textbf{3}, 346--350
  (2009).

\end{thebibliography}
\end{document}